\documentstyle[11pt,aaspp4,psfig]{article}  

\begin{document}

\title{Eclipsing Binaries in the OGLE Variable Star Catalog.~III.\\ 
Long-Period Contact Systems}

\author{\sc Slavek M. Rucinski}

\affil{Canada -- France -- Hawaii Telescope Co.\\
P.O.\ Box 1597, Kamuela, HI 96743\\
Electronic-mail: {\it rucinski@cfht.hawaii.edu\/}}

\bigskip
\centerline{\today}
 
\bigskip

\begin{abstract}
A sample of contact binaries discovered by the OGLE project
in Baade's Window, with orbital periods longer than one day 
and with available color and light-curve data, has been analyzed. 
It consists of only 32 systems, in contrast to 388 W~UMa-type 
systems with shorter periods which were analyzed before.
Most systems are very distant and are probably located close to
or in the galactic Bulge. Two groups of contact binaries are seen in 
the sample: (1)~a continuation of the W~UMa-type sequence, 
extending up to the orbital periods of 1.3 -- 1.5 day, but rather
sharply ending in this period range; (2)~an inhomogeneous group 
of rare systems with long periods up to 26 days, all with red colors
and relatively shallow eclipses. While the systems of the 
first group share most of the characteristics of the typical 
W~UMa-type systems (except that they are on the average brighter
and more distant, hence more reddened), the long-period 
systems do not seem to form an early-type extension of 
contact binaries, but may consist of a mixture of late-type objects, 
including tidally distorted red giants with invisible companions.
\end{abstract}

\section{INTRODUCTION}
\label{intro}

The microlensing projects provide, as their most important by-product, 
discoveries of large numbers of variable stars. The variable-star 
databases created by these projects have excellent statistical 
properties and permit addressing basic problems of stellar 
astrophysics, which require unbiased (or controlled bias) samples 
based on large numbers of objects. This paper is the third in a 
series devoted to eclipsing binaries discovered in the OGLE 
microlensing project in the direction of the Galactic Bulge, within 
the region of lower galactic extinction known as Baade's Window. 
The data used in this series come from the first three parts of 
the periodic variable-star catalogue covering fields BWC and 
BW1 to BW8 (Udalski et al. \markcite{uda94}1994, \markcite{uda95a}1995a, 
\markcite{uda95b}1995b)\footnote{The periodic variable star 
catalog is also available through the OGLE project Web site, via
{\it http://www.astrouw.edu.pl\/} or 
{\it http://www.astro.princeton.edu/$\sim$stanek/ogle\/}}. The fields 
BW9 to BW11 (Udalski et al. \markcite{uda96}1996) 
have not been used because no
data on extinction and reddening are available for these fields.

The first paper of the current series (Rucinski \markcite{ruc97a}1997a 
= R97a) addressed the average properties of the large numbers of the 
W~UMa-type contact-binary systems discovered 
by OGLE. It was shown that these binaries can be useful 
distance indicators along the line of sight all the way
to the Galactic Bulge at about 8 kpc, with this
line passing the center of the Galaxy 
at the distance of about 500 -- 600 pc. The W~UMa systems were 
seen to be uniformly distributed along this line, in agreement with
suggestions that, in their majority, they belong to an old 
galactic population. The analysis of the period and color distributions 
confirmed that the most common contact binaries belong to the 
population of the Old Disk (with a possible admixture of the Halo) 
Main Sequence turn-off  (MSTO) stars, within the well defined ranges
of the periods and colors: $0.25 < P < 0.65$ day and
$0.4 < V-I <1.4$. Their density was found to 
be relatively high, with the apparent (most probably about 2-times lower 
than the spatial) frequency of occurrence among the F -- K 
dwarfs of approximately one such a system per 
250 -- 300 main sequence stars.
 
The second paper of the series (Rucinski \markcite{ruc97b}1997b 
= R97b) analyzed properties of the light curves of the same systems using 
a simple, Fourier-analysis approach based on the lowest-term 
cosine coefficients. The light-curve amplitude distribution strongly 
suggested a mass-ratio distribution steeply climbing toward low 
mass-ratios (more unequal masses) indicating that the current 
sky-field sample (which contains predominantly large-amplitude 
systems) is heavily biased in favor of large mass-ratio ($q 
= M_2/M_1 \rightarrow 1$) systems. Contact system with unequal 
depths of eclipses (i.e. unequal temperatures of the components) 
appear only for orbital periods longer than 0.37 day, but their 
relative spatial frequency even above this cutoff is 
low, about 1/50 of the number 
of contact systems with comparable periods. The difference in depth 
of eclipses is correlated with the presence of unequally-high 
light-curve maxima. The most common type of the maxima asymmetry 
is in the sense of  the first (after the deeper eclipse) 
maximum being higher of the two. This indicates that 
the systems with asymmetric maxima may be semi-detached binaries with 
matter flowing from the hotter, more massive 
component and forming an accretion hot spot on the cooler component.

Papers R97a and R97b analyzed the eclipsing systems in the OGLE 
sample with orbital periods shorter than one day only. 
This paper analyzes the contact systems above the one-day 
period cutoff with a goal simply to see what sort 
of contact binaries exist at longer orbital periods.
There was also one specific reason to look at the long-period
binaries: As was pointed by 
Popper \markcite{pop82}(1982), 
some of the close, long-period, O-type, close 
binary systems show properties very similar to those of the 
genuine W~UMa-type systems. In particular -- and 
this may be taken as a definition of the class of contact binaries -- 
their components have sometimes equal temperatures in spite of 
unequal masses. Continuity of the properties of 
the contact-binary systems across the convective/radiative 
boundary at late-A/early-F spectral-type up to the O-type stars 
has never been proven, but its likelihood was pointed 
in numerous studies. Although very young stars are not expected to
exist in the OGLE sample due to the rapid emergence of the line of
sight from the Young Disk (for $b = -4^\circ$), at the distance
of about 2.5 to 3 kpc, the sample may still 
contain moderately young stars.

This paper is organized as follows: Section~\ref{types} discusses
selection of a sample of long-period contact systems, and
Sections~\ref{pc} and \ref{cmd} discuss the period--color and 
color--magnitude diagrams. The light curves of four systems with
periods longer than 1.5 days are described in Section~\ref{curves}.
Conclusions of the paper are given in Section~\ref{concl}. 
Papers R97a and R97b should be consulted for 
several details which have been omitted in this paper.

\section{THE ONE-DAY DIVIDING LINE AND
THE RESTRICTED LONG-PERIOD (RLP) SAMPLE}
\label{types}

By definition, the W~UMa-type binaries have periods shorter than one day.
On the other hand, the demarcation line at one day quite
commonly appears in the variable-star research and is 
obviously due to the night/day 
periodicity superimposed on any ground-based data. Is then the
one-day limit on the W~UMa-type periods a physical one?
In fact, contact systems with orbital periods of up to one day
were amply represented in the magnitude-limited sample 
in R97a and R97b, but the period distribution for a volume-limited 
sub-sample (R97a) did show a strong concentration in the period
range 0.25 to 0.65 day and a rapid drop for orbital periods 
longer than about 0.7 day. Therefore, the traditional definition of the
W~UMa-type is possibly not entirely artificial, but has
a deeper reason, which in R97a was attributed to
the evolution of close binary stars of old galactic population
in the final, Turn Off stage of the Main Sequence evolution. 
Here, we re-consider the selection of the contact-binary 
sample, with the use of the the Fourier analysis of the light 
curves, as in R97a and R97b, to see how
continuous is the transition in the numbers of the contact binaries
across the 1-day line.

\begin{figure}                                       
\centerline{\psfig{figure=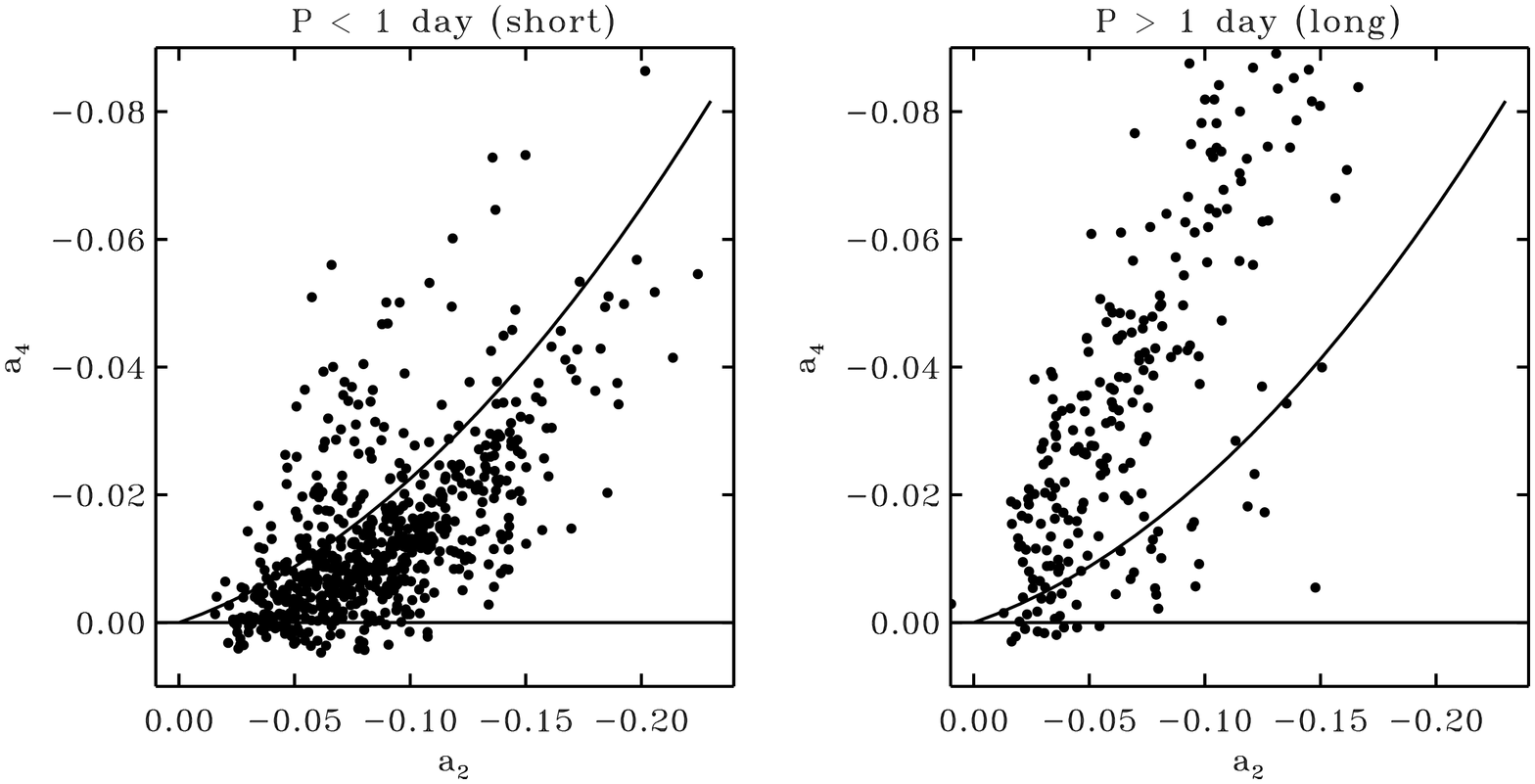,height=4.5in,angle=90}}
\vskip 0.5in
\caption{\label{fig1}
All 933 eclipsing systems of the OGLE sample are shown in 
two panels, separated by the orbital periods below and above one-day 
dividing line. Contact binaries analyzed in R97a, R97b and 
this paper have been selected by having the Fourier coefficients 
$a_2$ and $a_4$ below the inner-contact limiting curve 
$a_4^i$ (continuous line) and above the line $a_4 = 0$. The current paper 
analyses 32 long-period systems which fall into this domain,
and have available $V-I$ colors and 
well-defined light curves. This sample is called here 
RLP for Restricted Long Period.
}
\end{figure}

The full OGLE sample consists of 933 eclipsing binaries, 
among them 257 systems have orbital periods longer than one 
day. The Fourier coefficients $a_2$ and $a_4$ for all 
eclipsing systems are shown in Figure~\ref{fig1}. This figure 
contains also the curve which was used before to
divide the diagram into 
domains of contact (below) and non-contact (above) systems. 
This ``inner-contact'' curve, $a_4^i = a_2(0.125-a_2)$, was 
found in R97a on the basis of an extensive exploration 
of the whole parameter space characterizing light 
curves of contact binary systems, in terms of orbital inclinations, 
mass-ratios and fill-out factors (Rucinski \markcite{ruc93}1993 = R93). 
The calculations in R93 had an important limitation for the 
present context: They were all done for one set of atmospheric 
properties, in particular for one assumed gravity-brightening 
law, the so-called Lucy or convective-envelope law, 
$T_{eff} \propto g^{0.08}$. It is by no means obvious that 
the same inner-contact curve $a_4^i$ would apply to long period
systems if they happened to be hotter than the convective/radiative
envelope dividing line at spectral types around late-A/early-F.
Such hotter, radiative-envelope systems would presumably obey 
the von Zeipel law, $T_{eff} \propto g^{0.25}$.
In fact, as had been expected on the basis
of the particular spatial direction of the OGLE survey, 
all systems of the sample were found to be quite red.
However, without any spectroscopic data, we cannot exclude a remote
possibility that some of them are heavily reddened early-type
stars. For that reason, a new series of 
calculations with the assumed ``radiative'' gravity-brightening law
has been conducted. These calculations are briefly described 
in the Appendix to this paper, with the full results available 
electronically. They confirm that 
the inner-contact dividing-line established in R97a is approximately
valid also for the case of the radiative gravity-brightening law, 
and it tends to be only slightly too conservative at the large 
light-curve amplitude end. 

Figure~\ref{fig1} shows rather directly that 
the contact binaries dominate in numbers for the orbital
periods $P < 1$ day, 
with only a moderate number of detached systems above the 
inner-contact curve. In contrast, very few systems with 
$P > 1$ day fall into the contact-binary domain, between the 
inner-contact limiting curve $a_4^i$, and the line $a_4 = 0$ (from 
now on, this domain in the $a_2$, $a_4$ coefficient space 
will be called the Fourier filter). Thus, 
long-period contact systems are very rare. Because they 
must be intrinsically bright due to the large radiating surfaces, 
we should be able to see them to large distances. 
Therefore, the rarity of long-period contact binaries is even
more pronounced than may appear from the magnitude-limited OGLE
sample. Not knowing the discovery selection effects of the OGLE catalog 
for variable stars with periods comparable with durations of 
the OGLE observing seasons, we would be cautious to make any
statements for systems with periods of several weeks or months,
but the selection effects should be moderate and -- most
importantly -- similar for binaries with periods of one day to a few days.
Thus, the drop in numbers of contact systems around the period
of one day seems to be real. In the next section we suggest
that the normal sequence of contact systems extends to the periods
of 1.3 -- 1.5 day and then ends quite abruptly.

The sample used from this point on will be called RLP, for the 
Restricted Long Period sample. The 32 systems of this sample are
listed in Table~\ref{tab1}. It is an analogue of the 
R-sample used in R97a and R97b, and it is ``restricted'' in the sense
that it contains only 32 binaries 
which pass the Fourier filter from among 136 eclipsing long-period 
systems, have good light curves (overall quality of the 
``multiple-cosine, single-sine'' fit better than 0.04 
mag; cf. R97b) and have measured $V-I$ colors. However, 
12 among those 32 systems have small amplitudes of their 
light curves ($|a_2|< 0.05$) making any inferences susceptible 
to observational errors. In most cases, it is impossible to 
say anything about the light curves of those 12 systems; they 
would be normally called ellipsoidal variables, a name used in most 
cases to designate close binary systems with tidally strongly 
distorted components seen at low inclination angles and/or having 
very small mass-ratios. As we expand in Section~\ref{curves},
tidally-distorted single stars in systems with invisible companions
may also belong here.  

\section{THE PERIOD--COLOR DIAGRAM}
\label{pc}

The period--color (PC) relation is one of the most useful 
diagrams in studies of contact binaries. As was discussed in R97a, 
contact binaries cannot be bluer and/or cannot have shorter 
periods than a limiting line which was called the 
short-period blue envelope (SPBE). The SPBE has a meaning very 
similar to that of the Zero Age Main Sequence line on the 
color--magnitude diagram and offers us an important tool to judge 
the degree of evolution or reddening of a system. Both effects can 
shift a system away from this line: the reddening simply shifts 
in the color coordinate while evolution can lead to an increased 
period and a redder color. 

\begin{figure}                                       
\centerline{\psfig{figure=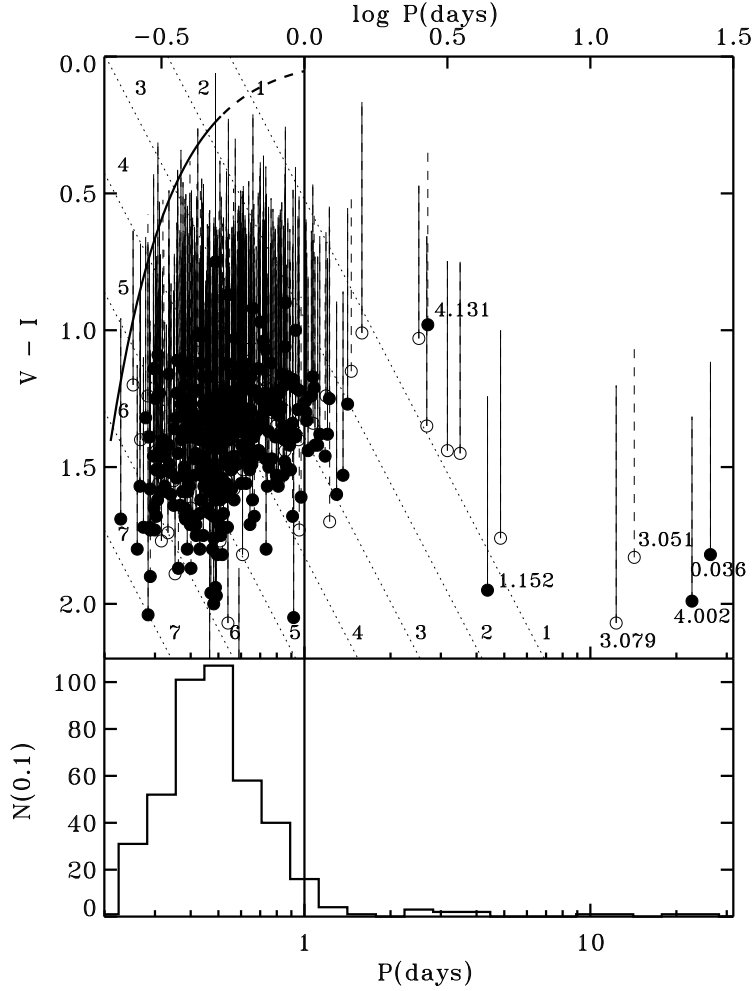,height=4.5in}}
\vskip 0.5in
\caption{\label{fig2}
The period--color diagram for the whole sample of contact 
binaries analyzed in this series of papers is shown in the upper
panel. The figure is 
divided into two parts by the one-day orbital period line. 
The filled circles are
used for systems showing light curve amplitudes larger than about 
$\Delta I \simeq 0.1$ (the strict criterion was 
$a_2 < -0.05$); the open circles are used for systems
with smaller amplitudes. The vertical lines pointing up from the 
circles give maximum reddening corrections, following Stanek 
(1996). 
The curve in the upper left corner is the short-period 
blue envelope, while the dotted slanted lines give loci of constant absolute 
magnitude $M_I$, see Figure 10 in R97a. The lower panel gives the
period histogram for the combined R and RLP samples. 
}
\end{figure}

Figure~\ref{fig2} shows the PC relation for all systems of the R-sample
(papers R97a and R97b) and the RLP-sample, together with the SPBE line
established for $P < 1$ day. The lower panel of the figure
gives the histogram of
the orbital periods for all contact binaries considered in this
series of papers. Figure~\ref{fig2} contains several interesting 
properties of the sample that we discuss in turn:
\begin{enumerate}
\item The conventional  
dividing line between the W~UMa and long-period contact binaries 
at one day is not obvious in the histogram of the orbital periods;
we also do not see any deficiency
in the one-day period bin which could be ascribed to discovery
selection effects in the OGLE data.
\item The RLP sample contains 21 systems in 
the 1 -- 2 day orbital period range, however
all but one have periods shorter than 1.5 day.
This abrupt drop in numbers of contact binaries at about 1.3 --
1.5 day strongly suggests that the class of
the genuine contact binaries of the W~UMa-type extends roughly
to this period. There are only 8 contact systems within the
$1.5 < P < 10$ day interval and 4 systems with $P > 10$ days.
\item The vertical, upward directed 
vectors in Figure~\ref{fig2} give the maximum 
values of reddening determined from interpolation in the maps 
of reddening and extinction by Stanek \markcite{sta96}(1996). These values 
were determined from data for the Bulge stars and may be 
over-estimates for nearby objects. However, they are probably 
reasonable approximations of the reddening for intrinsically 
luminous, long-period contact systems. Even after application of the 
relatively large reddening corrections, most of
the RLP sample systems 
remain surprisingly red, indicating that either they are in fact 
intrinsically red or that another source of reddening
increases their $V-I$ colors. On the PC diagram, these systems are 
located way below any reasonable extension of the SPBE. Thus, most systems, 
especially those with $P > 1.5$ days are not hot, Main-Sequence 
analogues of the W~UMa systems. 
\item Figure~\ref{fig2} shows lines of equal absolute magnitudes 
$M_I$, following the calibration in R97a: 
$M_I = -4.6\, \log P + 2.3\, (V-I)_0 - 0.2$. This calibration can 
only be used for binaries with fainter than absolute magnitude 
of about $M_I \simeq +1$ and for combinations of periods and 
colors encountered among the typical W~UMa-type systems. 
Therefore, it almost certainly does not apply to the long-period systems 
analyzed here. However, Figure~\ref{fig2} suggests that we can 
rather safely assume that, after reddening corrections are made, 
most of the RLP systems are brighter than $M_I = +1$, and perhaps 
by a large margin. For the extinction of $A_I \simeq 1.0$, systems 
located in or close to the Bulge at the distance of 8 kpc are expected 
to have $I \simeq 16.5$; this level is close to the median 
magnitude for the systems discussed in this paper (see 
Figure~\ref{fig3} below).
\item Among the 12 systems with $P > 1.5$ day, only four 
have variability amplitudes larger than $\Delta I \simeq 0.1$ mag
(formally, the criterion $a_2 < -0.05$ was used, with $a_2$ expressed
in intensity, not magnitude units). These are the
systems BW0.036, BW4.002, BW1.152 and 
BW4.131. These four systems\footnote{The naming convention used 
here is: BW for Baade's Window, followed by the OGLE field 
number, and then the variable number, after the dot. The central
field BWC is identified by zero.} are marked in 
Figure~\ref{fig2} and discussed further in the next sections.
\end{enumerate}

\section{THE COLOR--MAGNITUDE DIAGRAM}
\label{cmd}

\begin{figure}                                       
\centerline{\psfig{figure=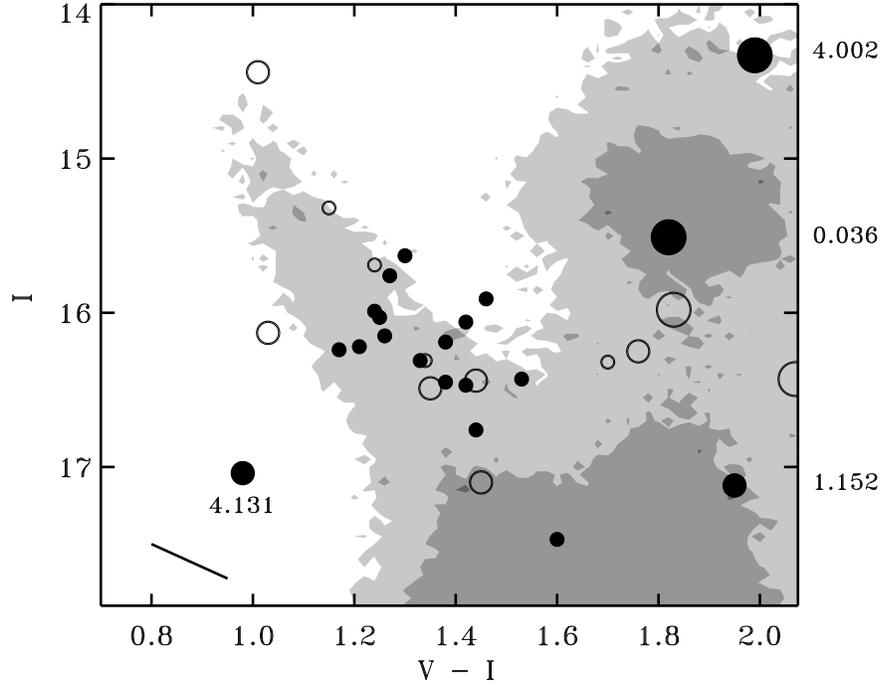,height=3.5in}}
\vskip 0.5in
\caption{\label{fig3}
The color--magnitude diagram for the systems of the RLP sample with 
sizes of the circles used to distinguish  three period groups 
among them: the small circles are for $1 < P < 1.5$ day, the medium 
size circles are for $1.5 < P < 10$ day, while the largest circles 
are for $P > 10$ day. As in other figures, filled circles mark 
systems with amplitudes larger than about 0.1 mag. Four systems which 
are described in Section~\ref{curves} are identified by their BW 
numbers. The gray contours give the density of the ordinary 
stars in Baade's Window, as in Figure~7 in R97a. Note that 
systems of the 1 -- 1.5 day group appear at the upper edge of the slanted 
sequence of old disk stars. The reddening vector is shown in the lower left
corner of the figure.
}
\end{figure}

Figure~\ref{fig3} shows the long-period contact systems on the 
observational color--magnitude diagram, together with the number 
density contours for Baade's Window stars, in the same way as 
in R97a. The orbital periods are coded by sizes of the circles in 
this figure. The open circles mark 12 systems with small 
variability amplitudes, measured here by the size of the 
second Fourier term, $|a_2| < 0.05$. Very little can be 
said about such binaries. In what follows, we will concentrate
on the 20 systems showing larger variability amplitudes, which are 
marked by filled circles.

Starting at the long-period end of the RLP sample with systems
having longer than 10 days: The brightest and reddest systems are
BW4.002 and BW0.036. The former, with the orbital period of 22.7 days, 
is located on the figure among bright red giants of the 
Bulge while the latter, with the period of 26.3 days, appears 
among stars of the Bulge Red Clump. Obviously, we have
no proof that these systems are at the distance of the Bulge;
they can be anywhere on the line of sight. But, because of the
long periods, there is no
question that both are red giants. If indeed contact systems, 
these would be extremely interesting objects demanding further 
studies. As was pointed by Eggleton \markcite{egg96}(1996),
only two red giants, UU~Cnc and 5~Cet, both with periods of about 96 days,
had been suspected to be contact red giants; Eggleton dismissed
this explanation for both binaries
suggesting semi-detached configurations instead.
The light curves of both systems are discussed in the next section.

The systems with periods between 1.5 and 10 days do not 
form a homogeneous group and do not 
show any clear association in terms of the accessible 
observational properties. Probably the most interesting is BW4.131, 
faint and thus probably distant, yet with a relatively blue apparent color. 

In contrast to the 1.5 -- 10 day period group, the contact 
systems with periods shorter than 1.5 days 
appear in a well defined location on the color--magnitude 
diagram in Figure~\ref{fig3}, slightly above the inclined band of the 
Main Sequence Turn Off (MSTO) stars of an old galactic population. 
The MSTO stars in Baade's Window are progressively reddened 
with distance so that their slanted sequence simulates the 
color-magnitude distribution of the relatively local, young-disk 
population (Kiraga et al. \markcite{kir97}1997); 
normally, in less reddened areas 
of the Milky Way, such stars form a band extending vertically 
over several magnitudes in brightness. Typical short-period W~UMa 
stars, which were analyzed in R97a, populate this band in the same 
way as other stars. As we can see in Figure~\ref{fig3}, the contact
systems in the $1 < P < 1.5$ day range are slightly differently
located and tend to populate the right upper edge of the 
old MSTO sequence. Thus, on the average, they are brighter 
and/or redder than typical MSTO (and, by implication, typical 
W~UMa-type) stars. Probably, it is a combination of several
factors why these systems delineate the red/bright edge of the 
MSTO and why the drop in their numbers at about 1.3 -- 1.5 day is 
so well defined. Most likely, these factors are:
(1)~a comparable, advanced age of the systems ending the MSTO 
evolution, (2)~a related to (1) limit on the mass for the old galactic 
population, and (3)~a drop in the spatial density of the disk 
stars as the line of sight goes past the distance of the Bulge. 
As we noted in the previous section, systems with $I \simeq 16.5$ 
are expected to be in the distance of the Bulge, 
provided their luminosities are higher than
corresponding to $M_I \simeq +1$, which is 
probably the case for most among contact systems with periods 
longer than one day. The systems of the 
1.0 -- 1.5 day group appear predominantly at $15.5 < I < 16.5$.
However, it should be kept in mind that rather large 
differences in extinction are observed within Baade's 
Window ($0.85 < A_I < 1.20$; Stanek \markcite{sta96}1996) 
so that the spread in $I$ is in fact surprisingly small. 
This can be taken as another indication of similar distances to these 
stars, probably in the space volume in the vicinity of the Bulge.

\section{THE LIGHT CURVES}
\label{curves}

\subsection{Relative depths of eclipses}

This paper attempts to preserve the spirit of the papers 
R97a and R97b in the sense that the analysis of the data is 
simple and straightforward. As in these papers, we use here only the 
Fourier coefficients $a_1$, $a_2$ and $a_4$. 

\begin{figure}                                       
\centerline{\psfig{figure=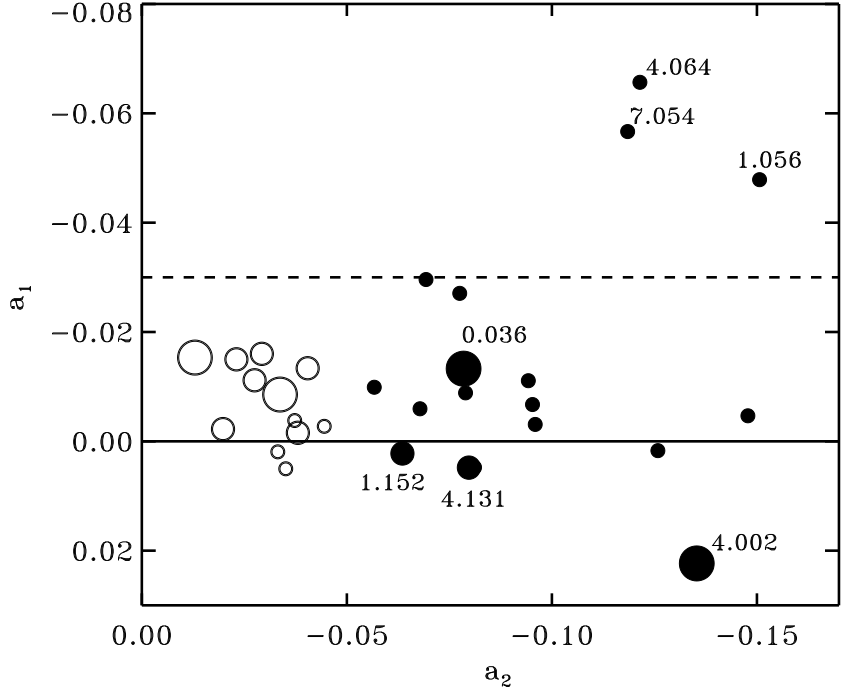,height=3.5in}}
\vskip 0.5in
\caption{\label{fig4}
The diagram $a_1$ versus $a_2$ which contains information  
on the difference of depths of eclipses. Only 3 systems 
(in the upper right corner) show moderately large 
depth differences which may indicate a poor 
thermal contact or lack of contact in a semi-detached configuration.
It is interesting that these three systems, 
BW1.056, BW4.064 and BW7.054, all have periods very close to one day,
whereas all systems with longer periods have almost equally deep
minima, as required by the definition of thermally stable contact
systems. Note that the $a_1$ coefficient for BW4.002 is positive,
as it should not be when the eclipses are correctly
identified and the phases are counted from the deeper one; 
the error results from the poor coverage of the primary minimum for this
star (see Figure~\ref{fig7}) and illustrates limitations of the
low-order Fourier coefficient approach. Open circles in this and
other figures are used for small-amplitude binaries showing $|a_2| < 0.05$.
}
\end{figure}

The first cosine coefficient $a_1$ can be used, as in R97b, to address 
the matter of the eclipse depth differences and the existence of
systems that are either not in thermal contact or perhaps
are semi-detached systems just mimicking good contact. Figure~\ref{fig4} 
shows the $a_1$ coefficient plotted versus $a_2$, the latter representing
the amplitudes of the light variations. As before, we 
use $a_1 = -0.03$ to delineate systems which have the 
eclipse depth differences too large to be considered as normal,
good-thermal-contact systems. We see only three such systems in the RLP 
sample, BW1.056, BW4.064 and BW7.054. Interestingly, all 
three have periods very close to one day: 1.057, 1.069 and 
1.009 day. All the remaining binaries
show light curves with sufficiently similar eclipses to be formally 
classified as good thermal and geometrical contact systems. The fraction
of 3 ``asymmetric'' systems among 20 systems with periods within
$1 < P <1.5$ day is somewhat high (but not statistically significant)
when compared with the percentages observed below one day (R97b).

\begin{figure}                                       
\centerline{\psfig{figure=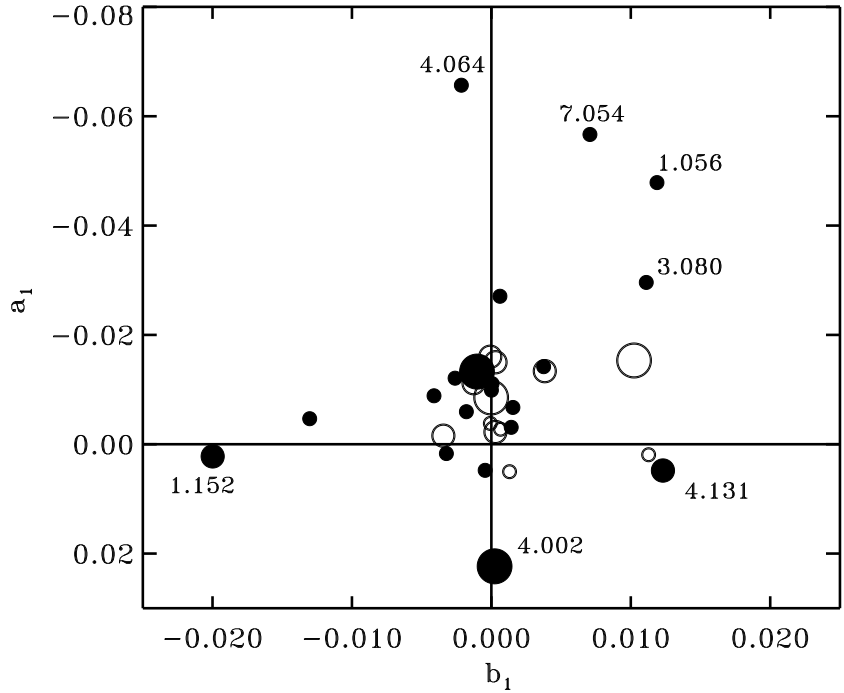,height=3.5in}}
\vskip 0.5in
\caption{\label{fig5}
Relation between the maxima asymmetry (measured by the first
sine coefficient $b_1$) and the depth difference (measured by the first
cosine coefficient $a_1$) for all systems of the RLP sample. The coding
of the symbols is the same as in the previous figure. Note that the sense
of the correlation is the same as in Figure~10 in R97b,i.e. the first
light-curve
maximum is higher for systems showing larger eclipse-depth differences.
}
\end{figure}

A correlation between depth of eclipses and light-curve
maxima asymmetry was found in R97b. The correlation was in the sense of the
larger depth difference going together with the higher first light
maximum (after the deeper eclipse). 
This correlation was driven by systems which
formally passed through the Fourier filter to be classified as contact
ones, yet showed eclipse depths disparate enough to suspect lack
of thermal contact or even semi-detached configurations. The
correlation could be most easily explained by an accretion region on
the side of the cooler, less-massive component. 
We see again the same correlation for the systems of the RLP sample
(Figure~\ref{fig5}), which is clearly visible mainly for the
systems with large eclipse differences mentioned above. 

\subsection{Light curve amplitudes}

As was shown in R97b, statistics of variability
amplitudes contains information on the mass-ratio ($Q(q)$)
distribution (in short: for $q \rightarrow 0$, only small
amplitudes are possible),
although extraction of this information would not be easy
since the latter enters through a convolution integral equation.

\begin{figure}                                       
\centerline{\psfig{figure=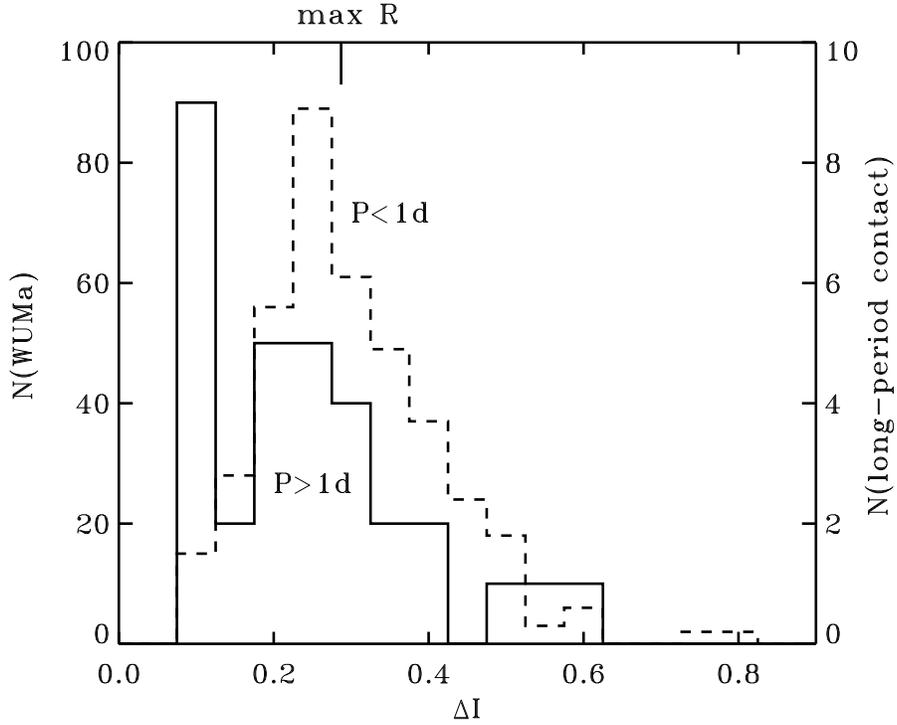,height=3.5in}}
\vskip 0.5in
\caption{\label{fig6}
The histogram of amplitudes for the RLP sample (continuous line
and right side vertical scale) 
is shown together with the histogram for the ordinary W~UMa-type 
systems of the R-sample, as in R97b (broken line and left side
vertical scale). The systems belonging to the isolated peak
at low amplitudes are shown by open circles in the
remaining figures of the paper. The mark at the
upper edge (``max R'') gives the largest possible amplitude for
a single, cool (convective) star filling its inner Roche 
equipotential in a system with $q=1$ and $i=90^\circ$ (see the
discussion of the system BW0.036 below). 
}
\end{figure}

With only 32 systems of the RLP sample, we can only ask a
simplified question: Is the amplitude distribution of the RLP systems
different from that for the genuine W~UMa systems of the R-sample?
Both distribution are shown in Figure~\ref{fig6}.
We see that, the small-number
statistics notwithstanding, the RLP sample consist of a
group of 9 small-amplitude systems, all
having $\Delta I \simeq 0.1$, and a larger group with the 
$\Delta I$-distribution which is somewhat similar to that for 
the typical W~UMa systems.
The small-amplitude end of the distribution must be
partly shaped by the OGLE discovery selection effects, but
we have no simple explanation for the relatively high
incidence of the small-amplitude systems in the sample forming
the isolated peak at $\Delta I \simeq 0.1$. It is possible that
it can be explained by the observational preference for such systems
in the OGLE sample, as they are all moderately bright with all, but one,
appearing at $I < 16.5$ mag. 

The rather rapid falloff from the side of large amplitudes
may indicate a mass-ratio distribution peaking at low values of $q$, 
as discussed in R97b for typical W~UMa-type systems. 
However, the peak for long period systems seems to be 
shifted toward smaller amplitudes which may indicate a tendency
for even smaller mass-ratios (more dissimilar stars) at longer
orbital periods. Most existing theoretical
scenarios envisage progressions to more extreme mass-ratios 
and to more massive and hotter primary components with advancement
of angular-momentum-loss driven evolution 
(for references, see Eggleton \markcite{egg96}1996).
These theoretical predictions have some support in the properties
of the most evolved among contact binaries. For example,
among 28 contact binaries discovered by Mazur et al.
\markcite{maz95}(1995) in the field of the open cluster Cr~261, 
the five which are located among the
Blue Stragglers of the cluster have systematically 
smaller amplitudes than the remaining systems (see Figure~14 in that 
paper). Also, Eggen and Iben \markcite{egg89}(1989) suggested that 
the small-amplitude system AW~UMa, the
well-known record holder in smallness of the mass-ratio among
W~UMa-type binaries, is a Blue Straggler in the local
volume close to the Sun. Thus,
while a link between the mass-ratio and the advancement of
evolution does seem to exist, do we see a similar 
link with the orbital period of a system? To answer this question,
the distribution shown in Figure~\ref{fig6} requires 
splitting into period bins.

\begin{figure}                                       
\centerline{\psfig{figure=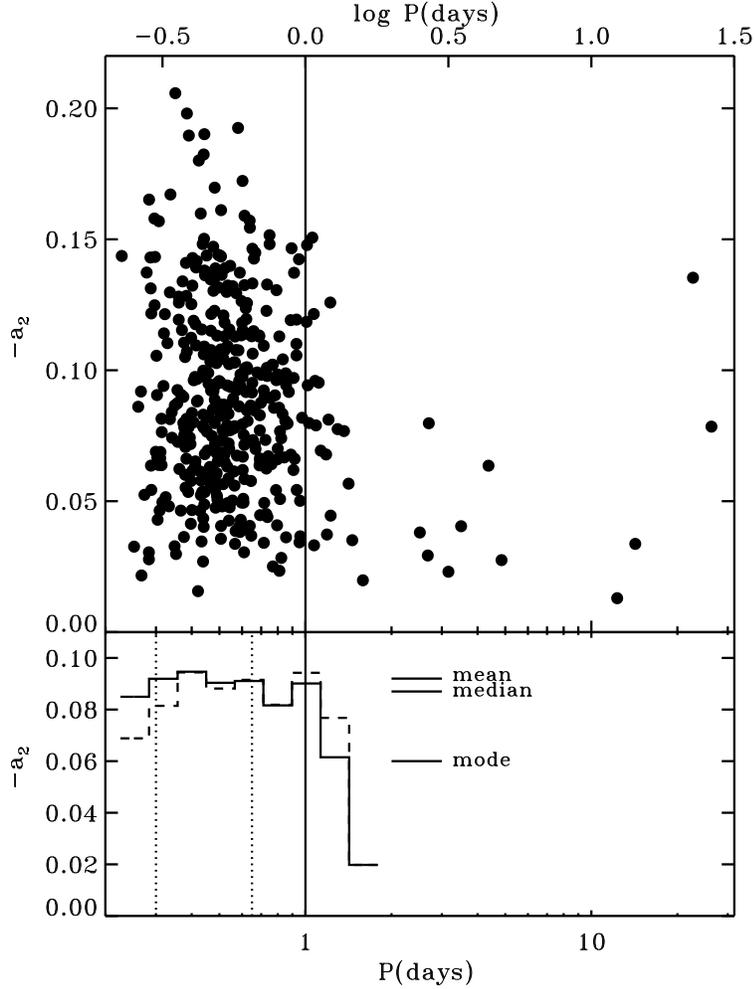,height=4.5in}}
\vskip 0.5in
\caption{\label{fig7}
The ``ellipsoidal'' components of the light curve variations, represented
by the second cosine coefficient $a_2$, 
and used here as measures of light-curve amplitudes,
are plotted versus the
orbital period in the upper panel. The lower panel gives the mean 
(continuous line) and median (broken line)
values in the period bins of $\Delta \log P = 0.1$. The marks at right
show the mean, median and modal values for the amplitude distribution
based on the short-period systems within $0.30 < P < 0.65$ day
interval which is delineated by two vertical dotted lines.
It is concluded that the apparent drop in
mean and median amplitudes seen at $P > 1$ day 
is probably due to the low number statistics (see the text).
}
\end{figure}

When analyzing the details of the light-curve amplitude
distributions, it is advantageous to use the coefficients 
$a_2$, instead of $\Delta I$. The latter  
scale\footnote{For small amplitudes, the relation is: 
$\Delta I \propto 2.2\, a_2$; for larger amplitudes the relation is
steeper: $a_2 = 0.1, 0.2, 0.3$, correspond to  
$\Delta I = 0.24, 0.55, 0.99$ mag.} with $a_2$, but offer 
an advantage of being free of the temperature differences
between components which affect the relative depths of the eclipses
and hence the values of $\Delta I$; the coefficients $a_2$ 
do retain the dependence on the mass-ratio and on the (assumed
randomly-distributed) orbital inclinations. Figure~\ref{fig7}
shows the values of $a_2$ for the combined short- and long-period
samples versus the orbital period. 
We see again that large amplitudes do not appear
among long-period systems. For typical W~UMa-type systems,
the mean and median values of
the amplitudes in bins of $\Delta \log P = 0.1$ 
are very similar across the whole period domain up to
one day. Then, beyond one day, the mean and median values appear to
fall down sharply. However, it is difficult to tell if this is simply
an effect of poorer statistics or a real effect as the mean and median
values of the amplitudes are expected to be biased. The reason is that
for progressively smaller sub-samples drawn from a skewed distribution,
the mean and the median values
will tend to drift toward the modal (most probable)
value of the distribution. If we use the amplitude
distribution for $0.3 < P < 0.65$ day range as a reference one,
we see that the mean and median values for the $1 < P <1.5$ day
indeed approach the modal value of that distribution 
(see the lower panel of Figure~\ref{fig7}). Thus, the apparent drop
in the mean and median values of the amplitudes for 
the group of the systems within $1 < P <1.5$ day
can be explained as due to the small number statistics. 
No statements on the global amplitude properties are possible for the 
long-period systems with $P > 1.5$ day.

\subsection{Long-period contact binary systems}

\begin{figure}                                       
\centerline{\psfig{figure=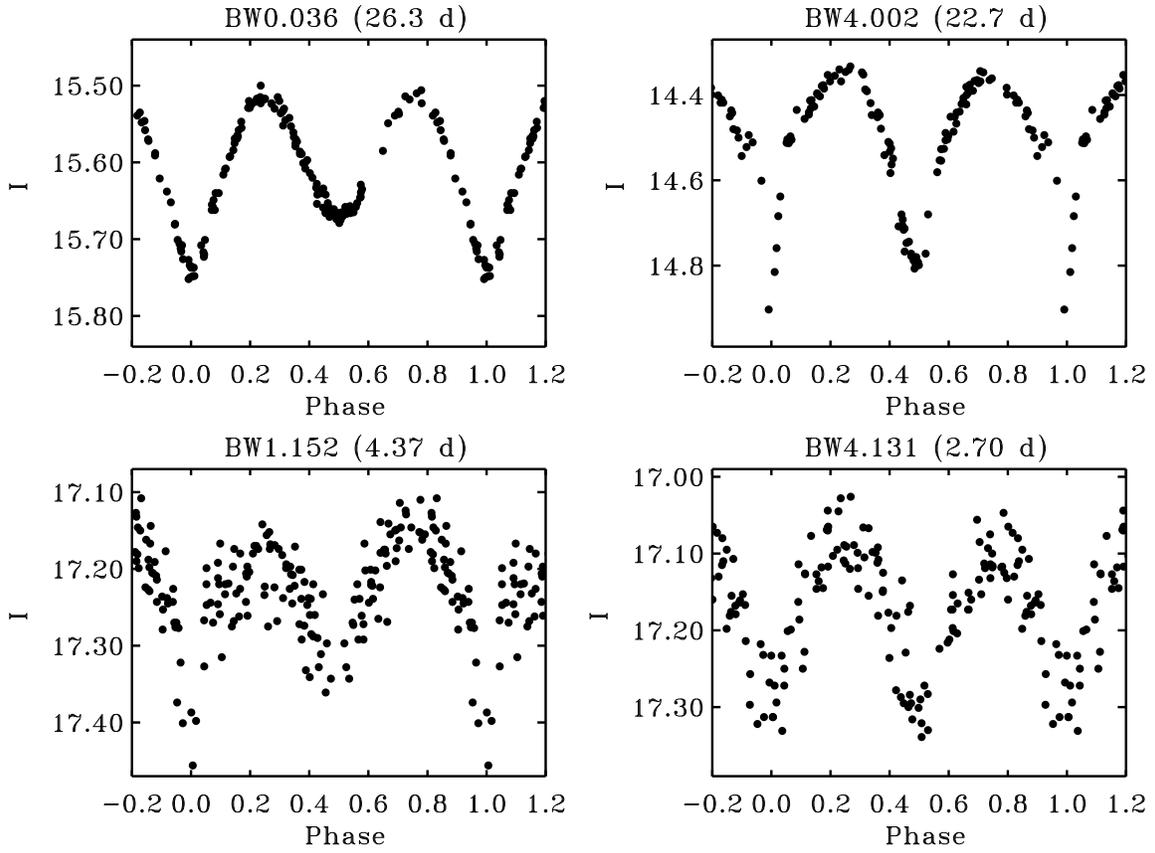,height=4.5in,angle=90}}
\vskip 0.5in
\caption{\label{fig8}
Light curves of four systems in the RLP sample with orbital 
periods longer than 1.5 days which have amplitudes larger than 0.1 mag. 
Only the light curve of BW0.036 is sufficiently well defined 
for any attempts of  modeling of it (see the next figure). The vertical
scales are the same in three panels and span 0.4 mag, except the one
for BW4.002 which spans 0.7 mag. 
}
\end{figure}

Probably the most interesting in this paper are the binaries 
which have passed the formal scrutiny to be included among 
contact systems and which have orbital periods longer 
than 1.5 days. Are these systems genuine contact binaries? 
Only four such systems have variability amplitudes larger 
than corresponding to  $|a_2| > 0.05$. The light curves for the four
systems are shown in Figure~\ref{fig8}. Only
BW4.002 shows $\Delta I > 0.5$ mag, but the deeper eclipse is
poorly defined in its light curve. Unfortunately, very little
can be said about these binaries because
light-curve solutions do not give unique results
for small variability amplitudes and without spectroscopic 
data on the mass-ratios. Here, additionally, the light curves for three of 
these systems are definitely too poor for any attempts 
at solving them. 
We are then left with only one long period system, BW0.036, 
which appears to be a contact one with a long period of 26.3 days,
but with a small amplitude of $\Delta I \simeq 0.24$ mag. 
Is it really a contact system? 
 
\begin{figure}                                       
\centerline{\psfig{figure=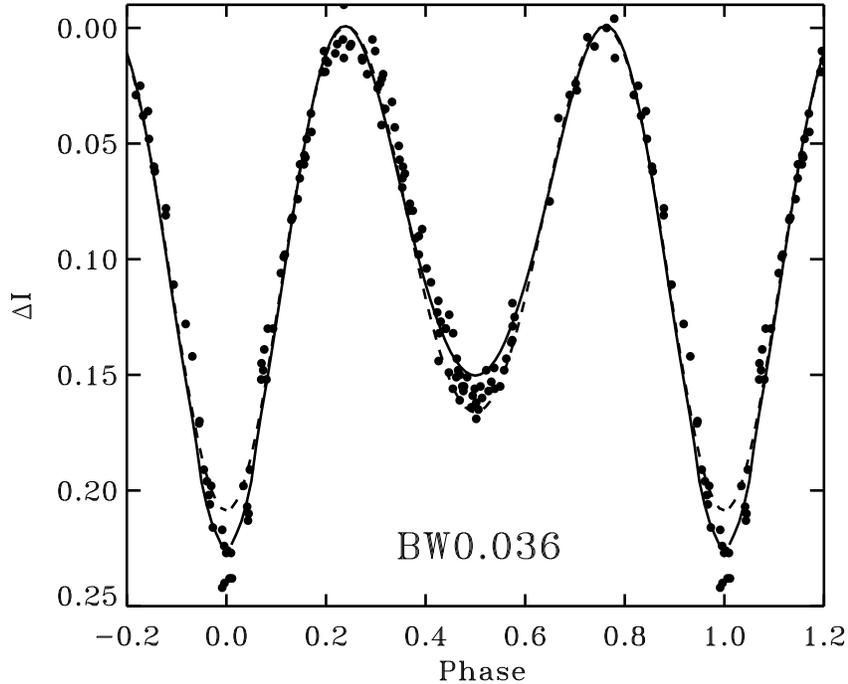,height=3.5in}}
\vskip 0.5in
\caption{\label{fig9}
The continuous line fit to the light curve of BW0.036 has been 
obtained by assuming that  the light variations are due 
to a single, convective-envelope star, almost filling its inner 
Roche lobe (smaller by 4 percent in radius) and 
tidally distorted by its same-mass ($q=1$) invisible companion.
The assumed orbital inclination was $i= 90^\circ$.
An experiment with a different,
radiative gravity-brightening exponent, $\beta = 0.25$,
in place of the convective one, $\beta = 0.08$ (in
$T_{eff} \propto g^\beta$), required $i=60^\circ$
(broken line); however, this was an internally inconsistent case
as the limb darkening $x = 0.65$
was kept the same, whereas for hot, radiative stars
$x$ would be substantially reduced, requiring the use
of a large orbital inclination. It is shown here only to illustrate
how non-unique, in terms of involved parameters, are light curves for
configurations such the postulated one for BW0.036. 
}
\end{figure}

If BW0.036 is a Red Clump giant in the Galactic Bulge, as its location
on the color--magnitude diagram would suggest (Figure~\ref{fig3}),
this would be a most unusual, highly evolved contact
system. However, it could be a
less distant star, a red giant or sub-giants, but there is little doubt
that it must be an evolved object.
We do not know any contact systems consisting of red giant stars
(cf. Section~\ref{cmd}), so we may consider
a different binary configuration: a single star, which is 
varying in brightness
because of the tidal distortion exerted by its much smaller companion.
A limited attempt at modeling the light curve of BW0.036 
following these lines has been made with the use 
of the light-curve synthesis program {\it Binary Maker 2\/} by 
Bradstreet \markcite{bra94}(1994). 
It was assumed that {\it the system is not a contact one, 
but that the only source of variability is an ellipsoidal 
variation of a tidally distorted single star, close to filling its Roche 
lobe, in a system containing two same-mass stars\/}. Its 
companion was assumed to be entirely invisible; it could be
a Main Sequence star or a collapsed object.
No solution was attempted, but only a ``proof-of-concept''
model. Thus, for simplicity,
the assumed parameters were: the mass-ratio $q=1$,
the inclination $i = 90^\circ$, the effective temperature
$T = 5000$ K, with the limb darkening coefficient $x = 0.65$
at $\lambda = 8500$ \AA\ 
and the gravity-brightening exponent $\beta = 0.08$, 
in $T_{eff} \propto g^\beta$. By varying only one parameter,
the surface potential, a reasonable fit was obtained with
$\Omega_1 = 3.85$ (Figure~\ref{fig9})\footnote{Dr.~Carla 
Maceroni very kindly made full, independent {\it solutions\/}
(not just model fits) 
of the light curves of BW0.036 using the Wilson-Devinney code; she
arrived at basically identical 
results.}. This potential corresponds to the star slightly 
smaller than the inner critical Roche lobe, by about 4 percent 
of its mean radius. Thus, it is feasible to explain a light curve
like that of BW0.036 by a non-contact geometry, although
the final check must await radial velocity analysis of the star.

The experiment described above does not prove that 
systems with equal-mass companions can explain all
cases of long-period systems which have light curves similar to
those of genuine contact binary stars. It simply shows
that small-amplitude light curves have very low information
content. Thus, not all binaries passing through the Fourier filter
must be contact systems. Still, we believe that 
systems such as BW0.036 are rare, and that it was discovered only
thanks to the tremendous number of red giants in the field.
Also, we note that
the amplitude distribution of contact binaries cannot be
strongly ``contaminated'' by systems like BW0.036 
as the latter could show only small amplitude variations: A limit
on the amplitude for the star filling its Roche lobe 
(for the limb and gravity laws as above) is reached for
$q=1$ and $i=90^\circ$ and is equal to $\Delta I \simeq 0.29$ mag or
$a_2 = -0.12$; the amplitudes decrease rapidly with departures 
from these limiting values of $q$ and $i$.
These considerations are also applicable to the amplitude
statistics of the genuine, short-period W~UMa-type contact binaries
which probably contains an admixture of red-dwarf/white-dwarf pairs 
such as V471~Tau. 

\section{CONCLUSIONS}
\label{concl}

This paper completes analysis of the contact systems in Baade's
Window on the basis of the OGLE sample by addressing the properties
of eclipsing binaries with periods longer than one day which passed
the Fourier light-curve shape filter. Only 32 such systems with
available color and light-curve data have
been identified; this should be compared with 388 contact binaries
with $P < 1$ day used in the previous two papers. The sample (RLP)
contains binaries of two types: (1)~those of a rather 
abruptly-ending extension of the W~UMa systems, with 
periods $P < 1.3 - 1.5$ day, and (2)~an inhomogeneous group
of systems with longer periods. 

The group of the contact binaries with the orbital periods
$P < 1.3 - 1.5$ day is well defined and its properties could be
understood in terms of the evolution of close binary stars of an
old galactic population. 
These stars are slightly redder and brighter than the sequence
of progressively reddened old Turn-Off stars, but almost certainly
belong to this sequence. The stars can be the most massive 
representatives of the population of old, close binary stars which
are entering the final stages of evolution, just before merging
of components, or can be analogues of contact Blue Stragglers
observed in globular clusters. Although mean and median
amplitudes of light variations for this group of systems are
systematically smaller than for the genuine W~UMa-type systems
with $P < 1$ day, this decrease may be entirely due to small
sample of the systems in the period range $1.0 < P <1.5$ day, coupled
with the skewed distribution of the amplitudes.

The second, long-period group was found to contain stars too red to form
a genuine continuation of the W~UMa-type sequence toward longer
periods. Analogues of the $\beta$ Lyrae semi-detached system or of 
the hot O-type contact binaries pointed out by Popper 
\markcite{pop82}(1982) were 
not really expected to show up in the sample due to the particular
galactic direction of the survey, and indeed none was found.
The systems seen in the OGLE sample are all quite
red suggesting configurations involving red
giants and sub-giants. However, in the 
absence of any spectroscopic information, we cannot 
exclude a possibility of an extremely large reddening in
some isolated cases of hot, distant stars. 
The simplest explanation of the long-period, red 
binaries would be in terms of spotted giants or, perhaps, binaries
with invisible (Main Sequence or collapsed)
companions. The first explanation can be ruled out
by the high stability of the light curves over the successive annual
OGLE seasons. Thus, basically for lack of any other alternatives, 
we are left with the binaries with only one component producing 
``ellipsoidal'' light variations.
Since almost all systems of the long-period group show small amplitudes,
this explanation appears to be valid; it certainly works in the case of 
the best-defined light curve of the star BW0.036. This system would be
then in a very special stage of the binary-star evolution and would be
visible only thanks to the multitude of red giants in Baade's Window.

\acknowledgments
Thanks are due to Carla Maceroni for her calculations of the light
curve fits for BW0.036 and to Carla Maceroni and
Bohdan Paczynski for very useful and constructive
comments and suggestions on the first version of this paper.

\appendix

\section{SIMPLE LIGHT CURVES COMPUTED FOR THE VON ZEIPEL'S LAW}

The set of simplified light curves computed in R93 was all 
calculated for solar-type stars with one assumed 
set gravity and limb darkening coefficients. As is 
well known, light-curves synthesis 
experiments applied to contact binaries always show
that the peculiar geometry of such 
systems -- rather than the atmospheric properties, entering 
through the limb and gravity coefficients and through the 
assumed emerging fluxes -- determines the shape of the light 
curves. Therefore, the results in R93 are expected to apply 
over a wide range of spectral types. However, the systems considered
in the current paper have been selected on the basis of their
long periods. Therefore, there is a chance that
some of them may be heavily reddened early-type stars.

A new set of light curves have been generated to test the 
sensitivity of the light curves, and especially  of the 
Fourier light-curve shape filter based on $a_2$ and
$a_4$, to the assumed atmospheric properties of 
the stars. As an extreme case, hot stars of approximately 
spectral type B0 (32,000~K), observed in the I-band (as
in the OGLE project) have been considered in this
auxiliary set of computations. Instead of the convective-envelope
(Lucy) gravity brightening law, $T_{eff} \propto g^{0.08}$, the
radiative-envelope (von Zeipel) law, $T_{eff} \propto g^{0.25}$,
was used. The additional
assumptions concerned the adopted bracketing
atmospheres (34,500~K and 30,250~K) which
were characterized by relative fluxes
1.078 and 0.922 and by the linear limb darkening 
coefficients 0.20 and 0.23.

The results of the computations, in the same format 
as the tables accompanying R93, will not be published, 
but are available through 
Internet\footnote{http://www.cfht.hawaii.edu/\~rucinski/}.
Of note are the following properties, which distinguish the
results for hot contact systems from those for the
solar-type case: (1)~Since the I-band is far red-ward of 
the spectrum peak for hot stars, the relative fluxes in the bracketing
atmospheres differ rather moderately; (2)~For the same reason,
the limb darkening coefficients are small; (3)~These two 
effects over-compensate the influence of the stronger
gravity brightening producing very similar, even
slightly less deep eclipses than for typical solar-type systems.

\begin{figure}                                       
\centerline{\psfig{figure=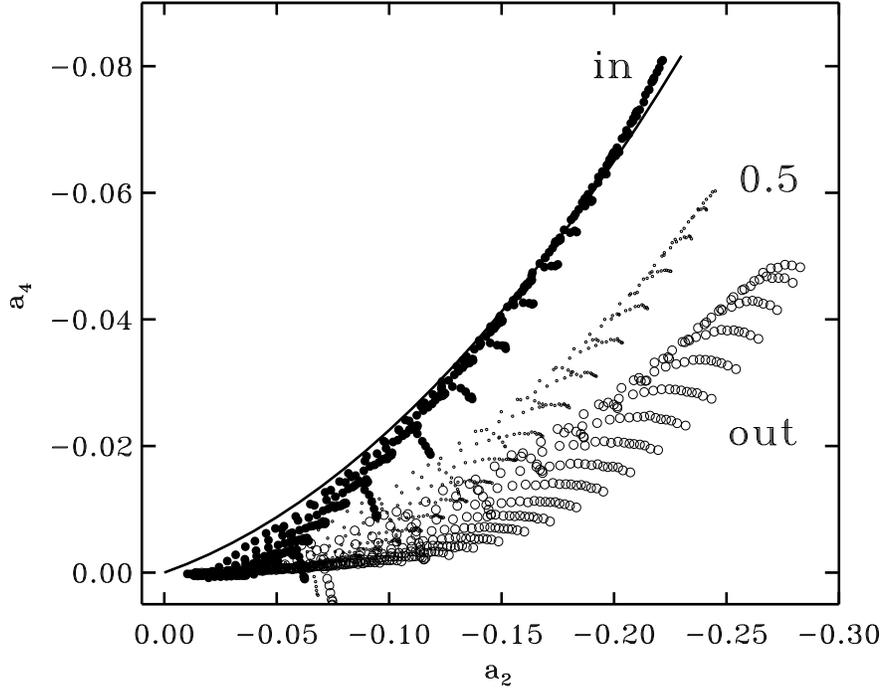,height=3.5in}}
\vskip 0.5in
\caption{\label{figA1}
Results of a new set of the contact binary light-curve computations for
the ``radiative'' (von Zeipel) gravity brightening law $T_{eff} \propto 
g^{0.25}$ are shown as a relation between the Fourier cosine coefficients
$a_2$ and $a_4$, as in Figure 6 in R93. The continuous line gives
the ``inner contact'' curve $a_4^i$ (see Section~\ref{types}) which
was used here and in R97a to select a sample of contact binary stars from
the OGLE survey data. 
}
\end{figure}

The diagram $a_4$ versus $a_4$, based on the new calculations,
is shown in Figure~\ref{figA1}.  This type of a diagram was
suggested in R93 as a simple tool for approximate 
estimates of the degree of contact and then used
in R97a for selection of the sample of contact binaries. 
As we can see in the figure,
the limiting ``inner-contact'' line, which was used in R97a,  
may serve quite well for the hot binaries. Detailed comparison 
of this figure and of Figure~6 in R93 reveals 
a few subtle differences, which can be ascribed primarily 
to the low values of the limb darkening coefficients for the
hot stars. The main change is that the light variations expected in 
the I-band for hot stars (measured here by $a_2$)
are actually slightly smaller than in 
the V-band for solar-type binaries, in spite of the use of the stronger 
(von Zeipel's) gravity-brightening law. 
This effect comes about because of the large 
areas of the projected disks seen during eclipses at low emergence 
angles, so that the reduced limb darkening 
in hot stars is more important, in fact over-compensates 
the stronger gravity brightening.

\newpage

\begin{deluxetable}{rrcccrrrr}
\tablewidth{0pc}
\tablecaption{\label{tab1} Long Period Contact Binary Systems}
\tablehead{
\colhead{BW}      &
\colhead{P}       &
\colhead{$I$}     &
\colhead{$V-I$}   &
\colhead{$\Delta I$} &
\multicolumn{4}{c}{Fourier coefficients} \nl
\colhead{number}  &
\colhead{(days)}  &
\multicolumn{3}{c}{} &
\colhead{cos~1$\phi$} &
\colhead{cos~2$\phi$} &
\colhead{cos~4$\phi$} &
\colhead{sin~1$\phi$}
}
\startdata
 0.036 &26.307 &15.51 & 1.82 & 0.24 &$-0.013$ &$-0.078$ &$-0.005$ &$-0.001$ \nl
 0.053 & 1.189 &15.69 & 1.24 & 0.13 &$-0.004$ &$-0.037$ &$-0.001$ &$-0.000$ \nl
 0.149 & 3.503 &17.10 & 1.45 & 0.21 &$-0.013$ &$-0.040$ &$-0.006$ &$ 0.004$ \nl
 1.056 & 1.057 &15.99 & 1.24 & 0.64 &$-0.048$ &$-0.151$ &$-0.040$ &$ 0.012$ \nl
 1.079 & 1.018 &16.31 & 1.33 & 0.27 &$-0.011$ &$-0.094$ &$-0.015$ &$ 0.000$ \nl
 1.152 & 4.368 &17.12 & 1.95 & 0.32 &$ 0.002$ &$-0.064$ &$-0.011$ &$-0.020$ \nl
 2.043 & 4.851 &16.25 & 1.76 & 0.11 &$-0.011$ &$-0.028$ &$-0.002$ &$-0.001$ \nl
 2.047 & 1.072 &16.31 & 1.34 & 0.17 &$ 0.002$ &$-0.033$ &$-0.004$ &$ 0.011$ \nl
 2.055 & 1.364 &16.43 & 1.53 & 0.29 &$-0.012$ &$-0.077$ &$-0.012$ &$-0.003$ \nl
 2.061 & 1.088 &16.47 & 1.42 & 0.30 &$-0.009$ &$-0.079$ &$-0.004$ &$-0.004$ \nl
 2.064 & 2.679 &16.49 & 1.35 & 0.12 &$-0.016$ &$-0.029$ &$-0.004$ &$-0.000$ \nl
 2.078 & 1.029 &16.76 & 1.44 & 0.30 &$-0.014$ &$-0.080$ &$-0.002$ &$ 0.004$ \nl
 3.023 & 1.013 &15.63 & 1.30 & 0.41 &$-0.005$ &$-0.148$ &$-0.005$ &$-0.013$ \nl
 3.051 &14.243 &15.98 & 1.83 & 0.13 &$-0.009$ &$-0.034$ &$-0.004$ &$-0.000$ \nl
 3.054 & 1.222 &16.03 & 1.25 & 0.36 &$ 0.002$ &$-0.126$ &$-0.017$ &$-0.003$ \nl
 3.070 & 1.225 &16.32 & 1.70 & 0.14 &$-0.003$ &$-0.044$ &$-0.003$ &$ 0.001$ \nl
 3.079 &12.313 &16.43 & 2.07 & 0.12 &$-0.015$ &$-0.013$ &$-0.001$ &$ 0.010$ \nl
 3.080 & 1.131 &16.45 & 1.38 & 0.25 &$-0.030$ &$-0.069$ &$-0.008$ &$ 0.011$ \nl
 4.002 &22.672 &14.33 & 1.99 & 0.57 &$ 0.022$ &$-0.135$ &$-0.034$ &$ 0.000$ \nl
 4.041 & 1.416 &15.76 & 1.27 & 0.21 &$-0.010$ &$-0.057$ &$-0.009$ &$ 0.000$ \nl
 4.064 & 1.069 &16.24 & 1.17 & 0.50 &$-0.066$ &$-0.121$ &$-0.023$ &$-0.002$ \nl
 4.082 & 3.163 &16.44 & 1.44 & 0.12 &$-0.015$ &$-0.023$ &$-0.001$ &$ 0.000$ \nl
 4.131 & 2.701 &17.04 & 0.98 & 0.31 &$ 0.005$ &$-0.080$ &$-0.014$ &$ 0.012$ \nl
 5.009 & 1.588 &14.44 & 1.01 & 0.11 &$-0.002$ &$-0.020$ &$-0.000$ &$ 0.000$ \nl
 5.138 & 1.296 &17.47 & 1.60 & 0.35 &$-0.027$ &$-0.078$ &$-0.013$ &$ 0.001$ \nl
 6.052 & 1.113 &16.06 & 1.42 & 0.27 &$-0.007$ &$-0.095$ &$-0.016$ &$ 0.002$ \nl
 6.059 & 1.204 &16.19 & 1.38 & 0.24 &$ 0.005$ &$-0.081$ &$-0.010$ &$-0.000$ \nl
 7.044 & 1.181 &15.91 & 1.46 & 0.21 &$-0.006$ &$-0.068$ &$-0.007$ &$-0.002$ \nl
 7.054 & 1.009 &16.15 & 1.26 & 0.43 &$-0.057$ &$-0.118$ &$-0.018$ &$ 0.007$ \nl
 7.057 & 1.081 &16.22 & 1.21 & 0.29 &$-0.003$ &$-0.096$ &$-0.006$ &$ 0.001$ \nl
 8.024 & 1.458 &15.32 & 1.15 & 0.14 &$ 0.005$ &$-0.035$ &$-0.001$ &$ 0.001$ \nl
 8.053 & 2.513 &16.13 & 1.03 & 0.19 &$-0.002$ &$-0.038$ &$-0.005$ &$-0.003$
\enddata
\end{deluxetable} 

\end{document}